\documentclass[twocolumn,prl,showkeys,preprintnumbers]{revtex4}%,nofootinbib]{revtex4}

\usepackage{amsmath}
\usepackage{amsfonts}
\usepackage[colorlinks,hyperindex]{hyperref}

\mathchardef\mhyphen="2D
\hypersetup{pdfpagemode=UseNone,pdfstartview=}%FitB}
\newcommand{\nc}{\newcommand}
\newcommand\fft[2]{\frac{#1}{#2}}
\newcommand\ft[2]{{\textstyle\frac{#1}{#2}}}
\newcommand\nn{{\nonumber}}
\newcommand{\beq}{\begin{equation}}
\newcommand{\eq}{\end{equation}}

\nc{\bea}{\begin{eqnarray}}
\nc{\ea}{\end{eqnarray}}
\nc{\be}{\begin{equation}}
\nc{\ee}{\end{equation}}
\nc{\barr}{\begin{array}}
\nc{\earr}{\end{array}}
\nc{\tr}{\text{tr}\,}

\makeindex

\begin{document}
%\preprint{MCTP-14-39}
%\preprint{ITP-UU-14/27}

\title{Central charges from the $\mathcal N=1$ superconformal index}

\author{Arash Arabi Ardehali}
\author{James T. Liu}
\affiliation{Michigan Center for Theoretical Physics, Randall Laboratory of Physics,\\
The University of Michigan, Ann Arbor, MI 48109--1040, USA}
\author{Phillip Szepietowski}
\affiliation{Institute for Theoretical Physics \& Spinoza Institute, \\
Utrecht University, 3508 TD Utrecht, The Netherlands}

\begin{abstract}
We present prescriptions for obtaining the central charges, $a$ and
$c$, of a four dimensional superconformal quantum field theory from
the superconformal index. At infinite $N$, for holographic theories dual
to Sasaki-Einstein 5-manifolds the prescriptions give the $\mathcal O(1)$
parts of the central charges. This allows us, among other things, to
show the exact AdS/CFT matching of $a$ and $c$ for arbitrary toric
quiver CFTs without adjoint matter that are dual to smooth
Sasaki-Einstein 5-manifolds. In addition, we include evidence
from non-holographic theories for the applicability of these results
outside of a holographic setting and away from the large-$N$ limit.
\end{abstract}


\maketitle

\section{Introduction}

Given a possibly strongly interacting quantum field theory, one of
the basic questions that can be asked is what are its degrees of
freedom.  In general, this appears to be a difficult problem.
However, with the addition of conformal symmetry, there is a growing
body of evidence that universal information on the spectrum of
operators is contained in the central charges.  In two dimensions,
this is evident from the Cardy formula \cite{Cardy:1986ie}, which
relates the asymptotic density of states to the central charge $c$,
as well as the Zamolodchikov $c$-theorem \cite{Zamolodchikov:1986gt}
governing flows between fixed points.  In four dimensions,
the central charges $a$ and $c$ control the entanglement entropy
\cite{Solodukhin:2008}, while the $a$-theorem
\cite{Komargodski:2011} suggests that $a$ is a proxy for the number
of degrees of freedom at conformal fixed points.

Additional support for a four-dimensional connection between central
charges and the spectrum comes from the recent observation
\cite{DiPietro:2014,Ardehali:2014} that the difference $c-a$ can be
obtained from the four-dimensional $\mathcal N=1$ superconformal
index \cite{Romelsberger:2005eg,Kinney:2005ej}.  This index counts
the number of shortened states in the spectrum, and for the
right-handed index is given by
\begin{equation}
\mathcal{I}^R
(t,y;a_i)=\mathrm{Tr}(-1)^{F}e^{-\beta\delta}t^{-2(E+j_{2})/3}
y^{2j_{1}}\prod a_i^{2s_i}, \label{eq:IRst}
\end{equation}
where $\delta=E-\fft32 r-2j_{2}$, and $\{E,j_1 ,j_2 ,r\}$ are the
quantum numbers of the superconformal group SU(2,2$|$1).  Here
$\beta$ regulates the infinite sum but otherwise drops out of the
index since only states with $\delta=0$ contribute. The final factor
above encodes global flavor symmetries with quantum numbers
$\{s_i\}$ and corresponding fugacities $\{a_i\}$. The left-handed
index $\mathcal I^L$ is similarly defined with the replacement
$r\to-r$ and $j_1\leftrightarrow j_2$.  The insertion of $(-1)^F$ is
what ensures that only the shortened spectrum contributes to the
index, and the result of \cite{DiPietro:2014,Ardehali:2014} is
consistent with the central charges $a$ and $c$ being among the
unrenormalized (protected) information in the theory
\cite{Anselmi:1997}.

There have been other attempts in the literature to relate the
central charges to the index.  In \cite{Razamat:2012}, the central
charge $c$ was noticed to play a role in the modular properties of
the $\mathcal{N}=2$ index, while in \cite{Buican:2014} a relation
was obtained for $2a-c$ of a CFT with $\mathcal{N}=2$ supersymmetry
(see also \cite{Shapere:2008}). Moreover, in \cite{Kim:2012ava} the
central charges were related to the so-called \emph{single-letter}
index, and in \cite{Assel:2014} it was observed that the central
charges dictate a specific relation between the supersymmetric
partition function on Hopf surfaces and the index. These results
suggest that it ought to be possible to obtain both of the central
charges $a$ and $c$ independently from the index.

In this letter we demonstrate that the superconformal index indeed
provides information about $a$ and $c$ separately.  This follows
from the recent work by Beccaria and Tseytlin \cite{Beccaria:2014}
that demonstrated that the one-loop corrections to $a$ and $c$ in
the holographic dual only receive contributions from the shortened
spectrum.  Since it is precisely this information that is captured
by the index, it is then possible to extract the corrections to $a$
and $c$ from the index. Following a similar approach as in
\cite{Ardehali:2014}, we find that the central charges are encoded
in the $t,y\rightarrow 1$ limit of the functions
%%%%%
\begin{eqnarray}
\hat{a}&=&\fft1{32}(t\partial_t+1)(-\ft92t\partial_t(t\partial_t+2)+\ft92(y\partial_y)^2-3)\hat
I(t,y),\nn\\
\hat{c}&=&\fft1{32}(t\partial_t+1)(-\ft92t\partial_t(t\partial_t+2)-\ft32(y\partial_y)^2-2)\hat
I(t,y),\nn\\
\label{eq:acIndex}
\end{eqnarray}
%%%%%
where $\hat I=(1-yt^{-1})(1-y^{-1}t^{-1})\mathcal I^+_{s.t.}$ is the
single-trace index with descendants removed and $\mathcal
I^+_{s.t.}\equiv\frac{1}{2}(\mathcal I^R_{s.t.}+ \mathcal
I^L_{s.t.})$. (The single-trace index is obtained from
Eq.~(\ref{eq:IRst}) by restricting the sum to the single-trace
spectrum and is natural from a holographic point of view.) The
fugacities are taken to one after acting with the differential
operator on $\hat I$ and the central charges are extracted as
%%%%
\begin{eqnarray}
a=\lim_{t\rightarrow1}\hat a(t,y=1),\quad c=\lim_{t\rightarrow1}
\hat c(t,y=1).\label{eq:acFromHats}
\end{eqnarray}
%%%%
Note also that the difference of these equations reproduces the
$c-a$ prescription of \cite{Ardehali:2014}.

Since Eq.~(\ref{eq:acIndex}) was derived from a one-loop computation in the
holographic dual, it only computes the subleading $\mathcal O(1)$ parts of $a$ and $c$
in holographic theories.  Curiously, however, it is possible to recover the full values of
$a$ and $c$ from these expressions for some classes of large-$N$ non-holographic
theories.  In any case, the result obtained from these equations may
be divergent when working in the large-$N$ limit, in which case the appropriate
prescription is to take the finite term in the Laurent expansions of $\hat a$
and $\hat c$ about $t=1$.

In order to highlight the potential divergences in $a$ and $c$, we
consider a series expansion of $\mathcal I^+_{s.t.}$, first around
$y=1$ and then around $t=1$.  Generically,
the expansion takes the form (see Sec.~IV of \cite{Ardehali:2014})
\begin{eqnarray}\label{eq:IndexExpanded}
\mathcal I_{s.t.}&=&\left(\fft{a_0}{t-1}+a_1+a_2(t-1)+\cdots\right)\nn\\
&&+(y-1)^2\left(\fft{b_0}{(t-1)^3}+\fft{b_1}{(t-1)^2}+\fft{b_2}{t-1}+\cdots\right)\nn\\
&&+\cdots.
\end{eqnarray}
We have dropped the $+$ superscript of $\mathcal I_{s.t.}$ assuming
that we are dealing with CP invariant theories; this will be the
running assumption in the rest of this letter. Applying Eq.~(\ref{eq:acIndex})
to this expression gives
\begin{eqnarray}\label{eq:cnaIndexExpanded}
\left.\hat a\right|_{y=1}&=&\fft{9(a_0-b_0)}{32(t-1)^2}-\fft{3(a_0+12a_2)-9(b_0-b_1+b_2)}{32}\nn\\
&&+\cdots,\nn\\
\left.\hat c\right|_{y=1}&=&-\fft{3(a_0-b_0)}{32(t-1)^2}-\fft{2(a_0+12a_2)+3(b_0-b_1+b_2)}{32}\nn\\
&&+\cdots.
\end{eqnarray}
Provided the single-trace index has the structure of (\ref{eq:IndexExpanded}), this
demonstrates that $\hat a$ and $\hat c$ have at most a double pole and no
single pole.  The prescription for removing the divergence then amounts to dropping
the double pole.

%%%%%%%%%%%%%%%%%%%%
\section{Large-$N$ theories with holographic dual}

We first examine the holographic case, since that is the framework
in which the expressions for $a$ and $c$ were derived.  More
specifically, we focus on four-dimensional SCFTs dual to IIB theory
on AdS$_5\times SE_5$
(and leave the study of other holographic settings to future work).
For these examples Eq.~(\ref{eq:acIndex}) gives only an $\mathcal O(1)$
subleading correction to the central charges, so in this section we expect to
only reproduce this subleading contribution which we denote by $\delta a$ and $\delta c.$
Of course, for such theories the computation of the $\mathcal O(1)$
part of $a$ and $c$ from the large-$N$ single-trace index of the
SCFT (or equivalently, the single-particle index of the gravity
side) follows directly from the work of Beccaria and Tseytlin
\cite{Beccaria:2014} on the one-loop contributions of bulk
one-particle states to the boundary central charges.  The use of the
superconformal index in Eq.~(\ref{eq:acIndex}) is at one level simply a
rewriting of the sum of the contributions over all bulk states.
However, the index does provide an alternative method for
regularizing the divergent sum over the KK towers in terms of
keeping the finite term in an expansion about $t=1$.

In principle, the application of Eq.~(\ref{eq:acIndex}) to a
holographic SCFT can also be viewed as a one-loop test of AdS/CFT.
In this sense, the result of \cite{Beccaria:2014} can be interpreted as a
test for the $\mathcal N=4$ theory, confirming and refining the
earlier results of \cite{Mansfield:2002pa,Mansfield:2003gs}. This
can be easily generalized to the case of arbitrary toric quiver CFTs
without adjoint matter that are dual to smooth Sasaki-Einstein
5-manifolds. The index of such a toric theory is \cite{Gadde:2010en,Eager:2012hx}
\begin{equation}
\mathcal{I}_{s.t.}=\sum_i
\frac{1}{t^{r_i/3}-1},\label{eq:toricIndex}
\end{equation}
where $r_i$ are the $R$-charges of extremal BPS mesons.
Applying (\ref{eq:acIndex}) to (\ref{eq:toricIndex}) gives
\begin{equation}
\hat a = -\frac{27}{32(t-1)^2}\sum_{i=1}^{n_v} \frac{1}{r_i} -
\frac{1}{32}\sum_{i=1}^{n_v}r_i+\cdots \label{eq:polToric}
\end{equation}
in an expansion about $t=1$. Keeping the finite
piece and noting that $\sum r_i=6(\mbox{\# nodes in the quiver})$,
we obtain
\begin{equation}
\delta a=-\fft3{16}(\mbox{\# nodes in the
quiver}).\label{eq:c-aToric}
\end{equation}
This matches the expected result for the $\mathcal O(1)$ part of $a$
based on the decoupling of a U(1) at each node in the quiver; since there are
no adjoint matter fields in the quiver, there are no additional
$\mathcal{O}(1)$ contributions to $a$ in the field theoretical
computation through $a=\frac{1}{32}(9\mathrm{Tr}R^3-3\mathrm{Tr}R)$.
The successful matching for the $\mathcal O(1)$ part of $c$ can now be
deduced either from a similar application of Eq.~(\ref{eq:acIndex})
to (\ref{eq:toricIndex}) or from the successful matching of $c-a$
reported in \cite{Ardehali:2014}.

We have also checked that Eq.~(\ref{eq:acIndex}) successfully
reproduces the $\mathcal O(1)$ part of the central charges of all
the other holographic theories discussed in \cite{Ardehali:2014}.
These include the $\mathcal{N}=4$ theory which has adjoint matter,
the singular $\mathbb{Z}_2$ orbifold, and the non-toric SPP and
del~Pezzo theories. (Of course, a test of AdS/CFT only occurs if the
index was computed on the gravity side. Since this was not the case
for the SPP and del~Pezzo theories, in these cases the matching is
only a confirmation of Eq.~(\ref{eq:acIndex}), and not a true test
of AdS/CFT.)

It is of course possible to perform a one-loop test by directly performing the
KK sum, and not going through the index as a regulator.  In particular, one
could proceed along the lines of
\cite{Ardehali:2013gra,Ardehali:2013xla,Ardehali:2013xya} by introducing a
$z^p$ regulator where $p$ is the KK level, and then taking the limit $z\to1$.
This type of regulator was recently justified for the $\mathcal N = 4$ theory in
\cite{Beccaria:2014} in terms of the ten-dimensional spectral $\zeta$-function,
and we have verified that it continues to provide a successful $\mathcal O(1)$
matching of both $a$ and $c$ for all the $\mathcal N=1$ cases discussed in
\cite{Ardehali:2013gra,Ardehali:2013xla,Ardehali:2013xya}.

\subsection{The second order pole in $a$ and $c$}

As seen in (\ref{eq:cnaIndexExpanded}), the coefficients of the
second order pole in $\hat a$ and $\hat c$, and hence $\hat c-\hat
a$ are all determined by the combination $a_0-b_0$. A relation was
given in \cite{Ardehali:2014} for the pole in $\hat c-\hat a$ in
terms of curvature invariants of the dual geometry.  Therefore
similar relations may be obtained for the coefficients of the pole
terms that Eq.~(\ref{eq:acIndex}) gives for $\hat a$ and $\hat c$ of
a holographic SCFT. The relation proposed in \cite{Ardehali:2014}
implies a negative coefficient for the pole in $\hat c-\hat a$, and
hence a positive one for $\hat a$ and a negative one for $\hat c$.

Because of the universal behavior of the second order pole, for all
SCFTs dual to IIB theory on AdS$_5\times SE_5$ the combination
\begin{equation}\label{eq:3c+a}
3\hat c+\hat a =
-\fft9{32}(t\partial_t+1)(2t\partial_t(t\partial_t+2)+1)\hat I(t,y),
\end{equation}
is finite at $t=1$.  In fact, assuming the expansion (\ref{eq:IndexExpanded})
for the index, this combination is always finite.
The finiteness can be traced to the absence of the $y$-dependent
operator in (\ref{eq:3c+a}). This particular combination of $a$ and
$c$ has been shown to be proportional to a supersymmetric Casimir
energy in \cite{Kim:2012ava} and further discussed and argued to be
regularization scheme independent in
\cite{Assel:2014,Assel:2014tba}. Here we find explicit evidence for
these statements of scheme independence. In particular, we see that
this quantity receives no contributions from states with arbitrarily
large dimension in the large-$N$ limit which would give rise to the
second order pole in $\hat a$ and $\hat c$ individually.
It would be interesting to understand this behavior
more completely.

%%%%%%%%%%%%%%%%%%%%
\section{Non-holographic SCFTs}

Although the expression (\ref{eq:acIndex}) was derived from a
holographic computation of the $\mathcal{O}(1)$ contributions to $a$
and $c$, we can nevertheless ask whether it can apply to
non-holographic SCFTs as well. Since the single-trace index is
inherently a large-$N$ construct, we start the discussion with
large-$N$ SCFTs.

Our primary example are the $A_k$ theories, the simplest of
which has $k=1$; this is SQCD without adjoint matter. The
single-trace index can be obtained in the Veneziano limit \cite{Dolan:2008,Ardehali:2014},
and application of Eq.~(\ref{eq:acIndex}) then gives
%%%%%
\begin{widetext}
%%%%%
%
\begin{eqnarray}
\hat a&=&\frac{9(2k-1)(k+1)}{128k(t-1)^2}+ \frac{-3+3k-12k^2+N_c^2(6+3k+15k^2)-36N_c^4/N_f^2}{8(k+1)^3}+\cdots,\nn\\
\hat c&=&-\frac{3(2k-1)(k+1)}{128k(t-1)^2}+
\frac{-2+5k-11k^2+N_c^2(7+5k+16k^2)-36N_c^4/N_f^2}{8(k+1)^3}+\cdots,
\end{eqnarray}
%%%%%
\end{widetext}
%%%%%
with the finite terms giving the full values of $a$ and $c$, and not just
their $\mathcal O(1)$ components \cite{Kutasov:1995np}.

This example demonstrates that the divergence at $t=1$ remains,
presumably as a large-$N$ effect, regardless of holography.  However
here the finite term recovers the full $\mathcal O(N^2)$ values of
both $a$ and $c$ in contrast with the holographic examples where the
result only gave the $\mathcal{O}(1)$ contributions. The difference
presumably lies in the type of large-$N$ limit taken. For the $A_k$
theories we have taken the Veneziano limit, i.e. $N_c \gg 1$ with
$N_c/N_f$ fixed. To emphasize one reason why this is different from
a holographic 't~Hooft limit, a distinction should be made in the
number of types (or flavors) of single-trace operators. In the
holographic setting this corresponds to the number of Kaluza-Klein
towers that exist in the reduction so we will refer to each flavor
as an individual tower. In the Veneziano limit there are an infinite
number of towers of single-trace operators, as opposed to a finite
number of towers arising in the holographic examples. This feature
is presumably what allows the index to capture the full expressions
for $a$ and $c$, including the $N^2$ terms.

\subsection{Moving away from large-$N$}

Since the index is well-defined even away from the large-$N$ limit,
Eq.~(\ref{eq:acIndex}) ought to be applicable to finite-$N$ theories
as well.  However, in this case the single-trace index is not well
defined, and a natural choice is to replace it by the plethystic log
\cite{Benvenuti:2006} of the full index.

Obtaining tractable analytic expressions for the superconformal
index of interacting theories at finite $N$ is generally difficult,
so instead we first comment on the general structure.  Following
\cite{Ardehali:2014}, we assume that the result of the plethystic
log gives a reduced index $\hat I$ that is a regular function with a
first order zero at $t=1$ when $y=1$. In this case, one can Taylor
expand around $t=y=1$
\begin{eqnarray}
\hat I(t,y) &=& f_1(t-1) + f_2(t-1)^2 + f_3(t-1)^3 + \cdots\nn\\
&& + (y-1)^2 \big(g_0 + g_1(t-1) + \cdots\big) + \cdots,\quad
\label{eq:finiteNguess}
\end{eqnarray}
where we have kept only the terms relevant for the calculation of
$\hat a$ and $\hat c$ in (\ref{eq:cnaIndexExpanded}). Comparison with
(\ref{eq:IndexExpanded}) then yields $a_0=b_0=f_1$.  Examination
of (\ref{eq:cnaIndexExpanded}) then demonstrates that the expressions
for $\hat a$ and $\hat c$ in (\ref{eq:acIndex}) remain finite.

Consistency with the result of Di~Pietro and Komargodski
\cite{DiPietro:2014} then gives $a_0=b_0=32(c-a)$, along with a
further condition $g_0+g_1=0$ that was obtained in
\cite{Ardehali:2014}. Combining this information with
(\ref{eq:cnaIndexExpanded}) we find
\begin{eqnarray}\label{eq:cnaIndexExpandedFiniteN}
\hat a&=&-\fft3{32}(a_0+9a_2) + \mathcal O(t-1),\nn\\
\hat c&=&-\fft1{32}(2a_0+27a_2) + \mathcal O(t-1).
\end{eqnarray}
Remarkably, only the $a_0$ and $a_2$ coefficients enter the
expressions for $a$ and $c$. This means, in particular, that the
central charges can be obtained from $\mathcal
I^{\text{finite-}N}_{s.t.}(t,1)$ where $y$ is set to unity.  Since
$y=1$ corresponds to a computation of the supersymmetric partition
function on the round $S^3$
\cite{Hama:2011ea,Dolan:2011rp,Gadde:2011ia}, we see that no
squashing is needed to have $a$ and $c$ separately encoded. One can
also turn Eq.~(\ref{eq:cnaIndexExpandedFiniteN}) around and write it
as an expansion of the finite-$N$ single-trace index. The result is
\begin{eqnarray}\label{eq:IndexExpandedFiniteN}
\mathcal
I^{\text{finite-}N}_{s.t.}(t,1)&=&\frac{32(c-a)}{t-1}+a_1-\frac{32}{27}(3c-2a)(t-1)\nn\\
&&+\cdots.
\end{eqnarray}
Note that the Hofman-Maldacena bound $3c\ge2a$ \cite{Hofman:2008}
guarantees that the coefficient of $t-1$ in the above expansion is
never positive.

As an example, consider the family of theories with a U(1)$^N$ gauge
group and $N_{\chi}$ neutral chiral multiplets (along with their
conjugates) having $R$-charges $R_i$. This class includes the magnetic
dual description of SQCD with $N_f=N_c+1$. The index is
\begin{eqnarray}
\mathcal{I}^{\text{finite-}N}_{s.t.}&=&N\left(1-\frac{1-t^{-2}}{(1-t^{-1} y)(1-t^{-1}
y^{-1})}\right)\nn\\
&&+\sum_{i=1}^{N_{\chi}}\frac{t^{-R_i}-t^{R_i-2}}{(1-t^{-1}
y)(1-t^{-1} y^{-1})}.\label{eq:U(1)Nindex}
\end{eqnarray}
Expanding $\mathcal{I}^{\text{finite-}N}_{s.t.}(t,1)$ around $t=1$ yields
\begin{eqnarray}
\mathcal{I}^{\text{finite-}N}_{s.t.}&=&\frac{-2(N+\sum
(R_i-1))}{t-1}-\sum
(R_i-1)\nn\\
&&-\frac{1}{3}\left(\sum(R_i-1)^3-\sum(R_i-1)\right)(t-1)\nn\\
&&+\cdots.
\label{eq:U(1)NindexExpanded}
\end{eqnarray}
Comparing with (\ref{eq:IndexExpanded}), it is now easy to see that
$a_0=-2\mathrm{Tr}R=32(c-a)$ and
$a_2=-\frac{1}{3}(\mathrm{Tr}R^3-\mathrm{Tr}R)=-\frac{32}{27}(3c-2a)$,
thus confirming (\ref{eq:IndexExpandedFiniteN}).

Taking the expression (\ref{eq:IndexExpandedFiniteN}) one step
further, we now consider the plethystic exponential of the
finite-$N$ single-trace index near $t=1.$ For this it is convenient
to define $t = e^{\beta}$ and expand near $\beta=0.$ From
(\ref{eq:IndexExpandedFiniteN}) we find
\begin{eqnarray}\label{eq:FullIndexExpandedFiniteN}
\mathcal
I^{\text{finite-}N}(e^{\beta},1)\kern-5em&&\nn\\
&=&\exp\left(\sum_{n=1}^\infty \frac{1}{n}\mathcal I^{\text{finite-}N}_{s.t.}(e^{n\beta},1) \right)\nn\\
&=&\exp\left(\sum_{n=1}^\infty\frac{32(c-a)}{n^2 \beta}+\frac{a'_1}{n}-\frac{8}{27}(3c+a)\beta +\cdots\right)\nn \\
&=& \exp\left(\frac{16\pi^2(c-a)}{3 \beta}+\frac{4}{27}(3c+a)\beta
+\cdots\right),
\end{eqnarray}
where in the final equality we have replaced the infinite sums on
$n$ with their $\zeta$-function regularized values and thrown away
the divergent harmonic series, i.e. %%
\begin{equation}
\sum_{n=1}^\infty \frac{1}{n^2} = \frac{\pi^2}{6},  \qquad \sum_{n=1}^\infty \frac{1}{n} \rightarrow 0, \qquad \sum_{n=1}^\infty 1 \rightarrow -\frac{1}{2}.
\end{equation}
%
%by discarding the harmonic series
Note that with this regularization we are neglecting potential
$\mathcal{O}(\beta^0)$ and $\mathcal{O}(\log\beta)$ terms in the
exponent of the index (see Eq.~(4.9) in \cite{DiPietro:2014} which
demonstrates the existence of such terms in the index of a free
vector multiplet). Nonetheless the $\mathcal{O}(1/\beta)$ and
$\mathcal{O}(\beta)$ terms in (\ref{eq:FullIndexExpandedFiniteN})
appear to be unambiguous. In particular, the leading behavior of the
result (\ref{eq:FullIndexExpandedFiniteN}) is consistent with the
generic results of \cite{DiPietro:2014} on supersymmetric partition
functions. Furthermore, the $\mathcal{O}(\beta)$ term in the
exponential of (\ref{eq:FullIndexExpandedFiniteN}) is precisely the
supersymmetric Casimir energy (\ref{eq:3c+a}) which was originally
obtained in \cite{Kim:2012ava,Assel:2014} from the single-letter
index.

The above discussion provides evidence that the expression (\ref{eq:acIndex}),
when applied to finite-$N$ theories, yields $a$ and $c$ directly, without any
needed subtraction.  With the assumption in (\ref{eq:finiteNguess}) on the form of the
single-trace index (whose validity is worth exploring), we can turn
our conjecture into one for the coefficient of the linear term in
the expansion of the single-trace index around $t=1$; this is shown
as the last term in Eq.~(\ref{eq:IndexExpandedFiniteN}). The fact
that this term reproduces the precise behavior in
\cite{Assel:2014,Kim:2012ava} provides a non-trivial test of this
statement.

At large-$N$, the expression for $\hat a$ and $\hat c$ formally
diverges at $t=1$. This divergence is related to the infinite sum
encountered when computing the single-trace index, which at
finite-$N$ would terminate at $\mathcal{O}(N)$ due to trace
identities. In the holographic examples, Eq.~(\ref{eq:acIndex})
computes the $\mathcal O(1)$ contribution to $a$ and $c$, while in
the $A_k$ theories in the Veneziano limit, it yields the complete
$\mathcal O(N^2)$ behavior.  The distinction between these two cases
appears to be related to the number of types (or flavors) of
single-trace operators present in the theory, with the holographic
cases having a finite $\mathcal O(1)$ number and the $A_k$ theories
having an infinite $\mathcal O(N^2)$ number.

It would be interesting to explore the pole structure and validity
of our prescriptions in (\ref{eq:acIndex}) and (\ref{eq:acFromHats})
for the indices of strongly coupled theories with a six dimensional
origin \cite{Gaiotto:2009we}. These theories have $\mathcal{O}(N^3)$
degrees of freedom and also admit a dual holographic description in
the large-$N$ limit \cite{Gaiotto:2009gz}. Results on the indices
are already available \cite{Gadde:2011,Gaiotto:2013}, although
their behavior near $t=1$ remains to be explored.

Finally, for the holographic examples, the leading
$\mathcal O(N^2)$ contributions to $a$ and $c$ are well understood
from the gravity dual in terms of the geometry of the
internal manifold \cite{Henningson:1998gx,Gubser:1998}. In the
field theory they appear in the behavior of the Hilbert series for
mesonic operators in the CFT \cite{Eager:2010yu}. Therefore, the
leading order central charges are encoded in the spectrum as well;
our results suggest that they are not, however, encoded in the
large-$N$ superconformal index. A proper understanding of this
distinction may shed light on the manifestation of a holographic
dual directly within field theory.

%%%%%%%%%%
\begin{acknowledgments}

JTL wishes to thank Cyril Closset for useful discussions. PS is grateful
to A.~Gadde and A.~Tseytlin for insightful email correspondence. We would also like to
thank L.~Di~Pietro  and Z.~Komargodski for useful comments.
This work is part of the D-ITP consortium, a program of the Netherlands
Organisation for Scientific Research (NWO) that is funded by the Dutch
Ministry of Education, Culture and Science (OCW), and is also supported
in part by the US Department of Energy under grant DE-SC0007859.

\end{acknowledgments}

%\bibliographystyle{unsrt}
%\bibliography{/home/thiago/bibtex/articles,/home/thiago/bibtex/books}

\end{document}